%% file: main.tex
\documentclass[sigconf,10pt,nonacm]{acmart}
\setcopyright{cc}
\setcctype{by-sa}

\input{preamble/preamble}

\makeatletter
\apptocmd{\@mkauthors}{%
  \global\setbox\mktitle@bx=\vbox{%
    \vskip 5mm%
    \unvbox\mktitle@bx%
    \vskip 5mm
  }%
}{}{\PackageError{spacing}{Could not patch author spacing}{}}
\makeatother

\begin{document}

\title[\system: a logic programming language to express and enforce network designs]{\system: a logic programming language\texorpdfstring{\\}{}to express and enforce network designs\\[2.178mm]} % The vspace is to make the text flush to bottom neatly.

\author{Tibor Schneider}
\affiliation{%
  \institution{ETH Zurich}%
  \city{Zurich}%
  \country{Switzerland}%
}
\author{Laurent Vanbever}
\affiliation{%
  \institution{ETH Zurich}%
  \city{Zurich}%
  \country{Switzerland}%
}

\Abstract{%
  Human errors remain the leading cause of network outages.
  To prevent them researchers propose general verifiers that analyze the network behavior.
  Yet, their (symbolic) models struggle to accurately reflect network devices, they fail to scale to large networks, and report misbehavior, not what configuration line caused it.
  We attribute all three challenges to specifications that describe the intended behavior but not \emph{how} operators intend to achieve it---the network's \emph{design}: which protocols to run and how to combine them.

  We present \system, a language for expressing network designs, and a checker that finds every violation and the responsible commands.
  Operators state their design decision as rules over configuration commands.
  \system then compiles them into a single Datalog program that checks the entire network directly on its configurations, not on the state they produce.
  Our prototype checks a 754-router network within seconds.
  Ultimately, a design enforced on every keystroke could give network operators the integrated development environment that programmers have long enjoyed.
}

\maketitle
\input{introduction}
\input{example}
\input{motivation}

\input{language}
\input{evaluation}
\input{related-work}
\input{discussion}

\paragraph{Ethical Concerns} This work raises no ethical concerns.

%\newpage
\bibliographystyle{ACM-Reference-Format} 
\bibliography{refs}

\end{document}

%% file: preamble/preamble.tex
\usepackage[utf8]{inputenc}
\usepackage{xspace}
\usepackage{enumitem}
\usepackage{cleveref}
\usepackage{float}

\usepackage{newfile}

\input{preamble/colors}
\input{preamble/listing}
\input{preamble/tikz}
\input{preamble/tikzmark}

\newoutputstream{abstractout}
\newcommand{\Abstract}[1]{%
  \openoutputfile{\jobname.abstract}{abstractout}%
  \addtostream{abstractout}{#1}%
  \closeoutputstream{abstractout}%
  \begin{abstract}%
    #1%
  \end{abstract}%
}

\newcommand{\system}{\textsc{Axiom}\xspace}
\newcommand{\extension}{\texttt{\small{.ax}}\xspace}
\newcommand{\networktoml}{\texttt{\small{network.toml}}\xspace}

%% file: preamble/colors.tex
\usepackage{xcolor}
\definecolor{slate50}{HTML}{f8fafc}
\definecolor{slate100}{HTML}{f1f5f9}
\definecolor{slate200}{HTML}{e2e8f0}
\definecolor{slate300}{HTML}{cbd5e1}
\definecolor{slate400}{HTML}{94a3b8}
\definecolor{slate500}{HTML}{64748b}
\definecolor{slate600}{HTML}{475569}
\definecolor{slate700}{HTML}{334155}
\definecolor{slate800}{HTML}{1e293b}
\definecolor{slate900}{HTML}{0f172a}
\definecolor{slate950}{HTML}{020617}
\definecolor{gray50}{HTML}{f9fafb}
\definecolor{gray100}{HTML}{f3f4f6}
\definecolor{gray200}{HTML}{e5e7eb}
\definecolor{gray300}{HTML}{d1d5db}
\definecolor{gray400}{HTML}{9ca3af}
\definecolor{gray500}{HTML}{6b7280}
\definecolor{gray600}{HTML}{4b5563}
\definecolor{gray700}{HTML}{374151}
\definecolor{gray800}{HTML}{1f2937}
\definecolor{gray900}{HTML}{111827}
\definecolor{gray950}{HTML}{030712}
\definecolor{zinc50}{HTML}{fafafa}
\definecolor{zinc100}{HTML}{f4f4f5}
\definecolor{zinc200}{HTML}{e4e4e7}
\definecolor{zinc300}{HTML}{d4d4d8}
\definecolor{zinc400}{HTML}{a1a1aa}
\definecolor{zinc500}{HTML}{71717a}
\definecolor{zinc600}{HTML}{52525b}
\definecolor{zinc700}{HTML}{3f3f46}
\definecolor{zinc800}{HTML}{27272a}
\definecolor{zinc900}{HTML}{18181b}
\definecolor{zinc950}{HTML}{09090b}
\definecolor{neutral50}{HTML}{fafafa}
\definecolor{neutral100}{HTML}{f5f5f5}
\definecolor{neutral200}{HTML}{e5e5e5}
\definecolor{neutral300}{HTML}{d4d4d4}
\definecolor{neutral400}{HTML}{a3a3a3}
\definecolor{neutral500}{HTML}{737373}
\definecolor{neutral600}{HTML}{525252}
\definecolor{neutral700}{HTML}{404040}
\definecolor{neutral800}{HTML}{262626}
\definecolor{neutral900}{HTML}{171717}
\definecolor{neutral950}{HTML}{0a0a0a}
\definecolor{stone50}{HTML}{fafaf9}
\definecolor{stone100}{HTML}{f5f5f4}
\definecolor{stone200}{HTML}{e7e5e4}
\definecolor{stone300}{HTML}{d6d3d1}
\definecolor{stone400}{HTML}{a8a29e}
\definecolor{stone500}{HTML}{78716c}
\definecolor{stone600}{HTML}{57534e}
\definecolor{stone700}{HTML}{44403c}
\definecolor{stone800}{HTML}{292524}
\definecolor{stone900}{HTML}{1c1917}
\definecolor{stone950}{HTML}{0c0a09}
\definecolor{red50}{HTML}{fef2f2}
\definecolor{red100}{HTML}{fee2e2}
\definecolor{red200}{HTML}{fecaca}
\definecolor{red300}{HTML}{fca5a5}
\definecolor{red400}{HTML}{f87171}
\definecolor{red500}{HTML}{ef4444}
\definecolor{red600}{HTML}{dc2626}
\definecolor{red700}{HTML}{b91c1c}
\definecolor{red800}{HTML}{991b1b}
\definecolor{red900}{HTML}{7f1d1d}
\definecolor{red950}{HTML}{450a0a}
\definecolor{orange50}{HTML}{fff7ed}
\definecolor{orange100}{HTML}{ffedd5}
\definecolor{orange200}{HTML}{fed7aa}
\definecolor{orange300}{HTML}{fdba74}
\definecolor{orange400}{HTML}{fb923c}
\definecolor{orange500}{HTML}{f97316}
\definecolor{orange600}{HTML}{ea580c}
\definecolor{orange700}{HTML}{c2410c}
\definecolor{orange800}{HTML}{9a3412}
\definecolor{orange900}{HTML}{7c2d12}
\definecolor{orange950}{HTML}{431407}
\definecolor{amber50}{HTML}{fffbeb}
\definecolor{amber100}{HTML}{fef3c7}
\definecolor{amber200}{HTML}{fde68a}
\definecolor{amber300}{HTML}{fcd34d}
\definecolor{amber400}{HTML}{fbbf24}
\definecolor{amber500}{HTML}{f59e0b}
\definecolor{amber600}{HTML}{d97706}
\definecolor{amber700}{HTML}{b45309}
\definecolor{amber800}{HTML}{92400e}
\definecolor{amber900}{HTML}{78350f}
\definecolor{amber950}{HTML}{451a03}
\definecolor{yellow50}{HTML}{fefce8}
\definecolor{yellow100}{HTML}{fef9c3}
\definecolor{yellow200}{HTML}{fef08a}
\definecolor{yellow300}{HTML}{fde047}
\definecolor{yellow400}{HTML}{facc15}
\definecolor{yellow500}{HTML}{eab308}
\definecolor{yellow600}{HTML}{ca8a04}
\definecolor{yellow700}{HTML}{a16207}
\definecolor{yellow800}{HTML}{854d0e}
\definecolor{yellow900}{HTML}{713f12}
\definecolor{yellow950}{HTML}{422006}
\definecolor{lime50}{HTML}{f7fee7}
\definecolor{lime100}{HTML}{ecfccb}
\definecolor{lime200}{HTML}{d9f99d}
\definecolor{lime300}{HTML}{bef264}
\definecolor{lime400}{HTML}{a3e635}
\definecolor{lime500}{HTML}{84cc16}
\definecolor{lime600}{HTML}{65a30d}
\definecolor{lime700}{HTML}{4d7c0f}
\definecolor{lime800}{HTML}{3f6212}
\definecolor{lime900}{HTML}{365314}
\definecolor{lime950}{HTML}{1a2e05}
\definecolor{green50}{HTML}{f0fdf4}
\definecolor{green100}{HTML}{dcfce7}
\definecolor{green200}{HTML}{bbf7d0}
\definecolor{green300}{HTML}{86efac}
\definecolor{green400}{HTML}{4ade80}
\definecolor{green500}{HTML}{22c55e}
\definecolor{green600}{HTML}{16a34a}
\definecolor{green700}{HTML}{15803d}
\definecolor{green800}{HTML}{166534}
\definecolor{green900}{HTML}{14532d}
\definecolor{green950}{HTML}{052e16}
\definecolor{emerald50}{HTML}{ecfdf5}
\definecolor{emerald100}{HTML}{d1fae5}
\definecolor{emerald200}{HTML}{a7f3d0}
\definecolor{emerald300}{HTML}{6ee7b7}
\definecolor{emerald400}{HTML}{34d399}
\definecolor{emerald500}{HTML}{10b981}
\definecolor{emerald600}{HTML}{059669}
\definecolor{emerald700}{HTML}{047857}
\definecolor{emerald800}{HTML}{065f46}
\definecolor{emerald900}{HTML}{064e3b}
\definecolor{emerald950}{HTML}{022c22}
\definecolor{teal50}{HTML}{f0fdfa}
\definecolor{teal100}{HTML}{ccfbf1}
\definecolor{teal200}{HTML}{99f6e4}
\definecolor{teal300}{HTML}{5eead4}
\definecolor{teal400}{HTML}{2dd4bf}
\definecolor{teal500}{HTML}{14b8a6}
\definecolor{teal600}{HTML}{0d9488}
\definecolor{teal700}{HTML}{0f766e}
\definecolor{teal800}{HTML}{115e59}
\definecolor{teal900}{HTML}{134e4a}
\definecolor{teal950}{HTML}{042f2e}
\definecolor{cyan50}{HTML}{ecfeff}
\definecolor{cyan100}{HTML}{cffafe}
\definecolor{cyan200}{HTML}{a5f3fc}
\definecolor{cyan300}{HTML}{67e8f9}
\definecolor{cyan400}{HTML}{22d3ee}
\definecolor{cyan500}{HTML}{06b6d4}
\definecolor{cyan600}{HTML}{0891b2}
\definecolor{cyan700}{HTML}{0e7490}
\definecolor{cyan800}{HTML}{155e75}
\definecolor{cyan900}{HTML}{164e63}
\definecolor{cyan950}{HTML}{083344}
\definecolor{sky50}{HTML}{f0f9ff}
\definecolor{sky100}{HTML}{e0f2fe}
\definecolor{sky200}{HTML}{bae6fd}
\definecolor{sky300}{HTML}{7dd3fc}
\definecolor{sky400}{HTML}{38bdf8}
\definecolor{sky500}{HTML}{0ea5e9}
\definecolor{sky600}{HTML}{0284c7}
\definecolor{sky700}{HTML}{0369a1}
\definecolor{sky800}{HTML}{075985}
\definecolor{sky900}{HTML}{0c4a6e}
\definecolor{sky950}{HTML}{082f49}
\definecolor{blue50}{HTML}{eff6ff}
\definecolor{blue100}{HTML}{dbeafe}
\definecolor{blue200}{HTML}{bfdbfe}
\definecolor{blue300}{HTML}{93c5fd}
\definecolor{blue400}{HTML}{60a5fa}
\definecolor{blue500}{HTML}{3b82f6}
\definecolor{blue600}{HTML}{2563eb}
\definecolor{blue700}{HTML}{1d4ed8}
\definecolor{blue800}{HTML}{1e40af}
\definecolor{blue900}{HTML}{1e3a8a}
\definecolor{blue950}{HTML}{172554}
\definecolor{indigo50}{HTML}{eef2ff}
\definecolor{indigo100}{HTML}{e0e7ff}
\definecolor{indigo200}{HTML}{c7d2fe}
\definecolor{indigo300}{HTML}{a5b4fc}
\definecolor{indigo400}{HTML}{818cf8}
\definecolor{indigo500}{HTML}{6366f1}
\definecolor{indigo600}{HTML}{4f46e5}
\definecolor{indigo700}{HTML}{4338ca}
\definecolor{indigo800}{HTML}{3730a3}
\definecolor{indigo900}{HTML}{312e81}
\definecolor{indigo950}{HTML}{1e1b4b}
\definecolor{violet50}{HTML}{f5f3ff}
\definecolor{violet100}{HTML}{ede9fe}
\definecolor{violet200}{HTML}{ddd6fe}
\definecolor{violet300}{HTML}{c4b5fd}
\definecolor{violet400}{HTML}{a78bfa}
\definecolor{violet500}{HTML}{8b5cf6}
\definecolor{violet600}{HTML}{7c3aed}
\definecolor{violet700}{HTML}{6d28d9}
\definecolor{violet800}{HTML}{5b21b6}
\definecolor{violet900}{HTML}{4c1d95}
\definecolor{violet950}{HTML}{2e1065}
\definecolor{purple50}{HTML}{faf5ff}
\definecolor{purple100}{HTML}{f3e8ff}
\definecolor{purple200}{HTML}{e9d5ff}
\definecolor{purple300}{HTML}{d8b4fe}
\definecolor{purple400}{HTML}{c084fc}
\definecolor{purple500}{HTML}{a855f7}
\definecolor{purple600}{HTML}{9333ea}
\definecolor{purple700}{HTML}{7e22ce}
\definecolor{purple800}{HTML}{6b21a8}
\definecolor{purple900}{HTML}{581c87}
\definecolor{purple950}{HTML}{3b0764}
\definecolor{fuchsia50}{HTML}{fdf4ff}
\definecolor{fuchsia100}{HTML}{fae8ff}
\definecolor{fuchsia200}{HTML}{f5d0fe}
\definecolor{fuchsia300}{HTML}{f0abfc}
\definecolor{fuchsia400}{HTML}{e879f9}
\definecolor{fuchsia500}{HTML}{d946ef}
\definecolor{fuchsia600}{HTML}{c026d3}
\definecolor{fuchsia700}{HTML}{a21caf}
\definecolor{fuchsia800}{HTML}{86198f}
\definecolor{fuchsia900}{HTML}{701a75}
\definecolor{fuchsia950}{HTML}{4a044e}
\definecolor{pink50}{HTML}{fdf2f8}
\definecolor{pink100}{HTML}{fce7f3}
\definecolor{pink200}{HTML}{fbcfe8}
\definecolor{pink300}{HTML}{f9a8d4}
\definecolor{pink400}{HTML}{f472b6}
\definecolor{pink500}{HTML}{ec4899}
\definecolor{pink600}{HTML}{db2777}
\definecolor{pink700}{HTML}{be185d}
\definecolor{pink800}{HTML}{9d174d}
\definecolor{pink900}{HTML}{831843}
\definecolor{pink950}{HTML}{500724}
\definecolor{rose50}{HTML}{fff1f2}
\definecolor{rose100}{HTML}{ffe4e6}
\definecolor{rose200}{HTML}{fecdd3}
\definecolor{rose300}{HTML}{fda4af}
\definecolor{rose400}{HTML}{fb7185}
\definecolor{rose500}{HTML}{f43f5e}
\definecolor{rose600}{HTML}{e11d48}
\definecolor{rose700}{HTML}{be123c}
\definecolor{rose800}{HTML}{9f1239}
\definecolor{rose900}{HTML}{881337}
\definecolor{rose950}{HTML}{4c0519}

%% file: preamble/listing.tex
\usepackage{listings}
\usepackage{pmboxdraw}
\usepackage{fancyvrb}

\newcommand{\textRel}{\color{violet800}}
\newcommand{\textStr}{\color{emerald700}}
\newcommand{\textVar}{\color{red800}}
\newcommand{\textOp}{\bfseries\color{blue800}}
\newcommand{\textTy}{\color{yellow700}}

\newcommand{\Rel}[1]{{\ttfamily\small\textRel#1}}
\newcommand{\Str}[1]{{\ttfamily\small\textStr#1}}
\newcommand{\Var}[1]{{\ttfamily\small\textVar#1}}
\newcommand{\Op}[1]{{\ttfamily\small\textOp#1}}
\newcommand{\Ty}[1]{{\ttfamily\small\textTy#1}}
\newcommand{\Code}[1]{{\ttfamily\small#1}}

\lstdefinelanguage{Axiom}{
  morekeywords={rel, def, set, assert, if, for, apply, deny, allow, doc, and, or, not},
  morekeywords=[2]{Int, Real, String, IpAddr, IpNet, MacAddr},
  morekeywords=[3]{self},
  morekeywords=[4]{Config, Interface, Link, Device},
  sensitive=true,
  morecomment=[l]{//},
  morestring=[b]",
}

\lstdefinestyle{code}{
  language=Axiom,
  basicstyle=\ttfamily\scriptsize,
  keywordstyle=\textOp,
  keywordstyle=[2]\textTy,
  keywordstyle=[3]\textOp,
  keywordstyle=[4]\textRel,
  commentstyle=\color{neutral500},
  stringstyle=\textStr,
  frame=none,
  numbers=left,
  numberstyle=\ttfamily\tiny\color{neutral400},
  numbersep=5pt,
  xleftmargin=1.65em,
  showstringspaces=false,
  columns=fullflexible,
  keepspaces=true,
  breaklines=false,
  aboveskip=0.5em, belowskip=0.5em,
  escapeinside={(*}{*)},
  moredelim=**[is][\btStartMark]{@}{@},
}

\crefname{lstnumber}{line}{lines}
\Crefname{lstnumber}{Line}{Lines}

\lstdefinelanguage{Cisco}{
  morekeywords={no},
  sensitive=true,
  morecomment=[l]{!},
  morestring=[b]",
  moredelim=**[is][\btStartMark]{@}{@},
}

%% file: preamble/tikz.tex
\usepackage{tikz}
\usetikzlibrary{calc, math, arrows.meta, positioning, shapes, fadings, patterns, fit, arrows}

\tikzfading[
  name=fade north partial,
  bottom color=transparent!0,   % fully opaque at bottom
  top color=transparent!90,     % 10% opaque at top (80% transparent)
]

\pgfdeclarelayer{background}
\pgfdeclarelayer{foreground}
\pgfsetlayers{background,main,foreground}

\usepackage{pgfplots}
\pgfplotsset{compat=1.18}

\pgfplotsset{
  keep if/.style args={#1 is #2}{
    filter discard warning=false,
    unbounded coords=discard, 
    x filter/.code={
      \edef\tempa{\thisrow{#1}}
      \edef\tempb{#2}
      \ifx\tempa\tempb
      \else
        
      \fi
    }
  },
  keep if both/.style args={#1 is #2 and #3 is #4}{
    filter discard warning=false,
    unbounded coords=discard, 
    x filter/.code={
      \edef\tempa{\thisrow{#1}}
      \edef\tempb{#2}
      \edef\tempc{\thisrow{#3}}
      \edef\tempd{#4}
      \ifx\tempa\tempb
        \ifx\tempc\tempd
        \else
          
        \fi
      \else
        
      \fi
    }
  },
  keep if all three/.style args={#1 is #2 and #3 is #4 and #5 is #6}{
    filter discard warning=false,
    unbounded coords=discard, 
    x filter/.code={
      \edef\tempa{\thisrow{#1}}
      \edef\tempb{#2}
      \edef\tempc{\thisrow{#3}}
      \edef\tempd{#4}
      \edef\tempe{\thisrow{#5}}
      \edef\tempf{#6}
      \ifx\tempa\tempb
        \ifx\tempc\tempd
          \ifx\tempe\tempf
          \else
            
          \fi
        \else
          
        \fi
      \else
        
      \fi
    }
  },
}

%% file: preamble/tikzmark.tex
\usetikzlibrary{tikzmark}

\newcounter{tikzlstmark}
\newlength{\tmheight}
\newlength{\tmdepth}
\AtBeginDocument{%
  \settoheight{\tmheight}{\ttfamily\scriptsize Ablk|}%
  \settodepth{\tmdepth}{\ttfamily\scriptsize gjpqy}%
}
\newcommand\btStartMark{%
  \stepcounter{tikzlstmark}%
  \tikzmark{lm-start-\thetikzlstmark}%
  \bgroup\aftergroup\btEndMark%
}
\newcommand\btEndMark{%
  \tikzmark{lm-end-\thetikzlstmark}%
  \egroup%
}
\newcommand\lmFit[2][]{%
  \tikz[overlay, remember picture]{%
    \coordinate[yshift=\tmheight] (lm-#2-topleft) at (pic cs:lm-start-#2);%
    \coordinate[yshift=-\tmdepth] (lm-#2-botright) at (pic cs:lm-end-#2);%
    \node[inner sep=1pt, fit={(lm-#2-topleft) (lm-#2-botright)}, #1] (lm-#2) {};%
  }%
}
\newcommand\lmfit[2][]{%
  \begin{tikzpicture}[overlay, remember picture]%
    \coordinate[yshift=\tmheight] (lm-#2-topleft) at (pic cs:lm-start-#2);%
    \coordinate (lm-#2-botright) at (pic cs:lm-end-#2);%
    \node[inner sep=1pt, fit={(lm-#2-topleft) (lm-#2-botright)}, #1] (lm-#2) {};%
  \end{tikzpicture}%
}

%% file: introduction.tex
\newcommand\cModel{\textit{(i)}}
\newcommand\cTime{\textit{(ii)}}
\newcommand\cFeedback{\textit{(iii)}}

\section{Introduction}
Network misconfigurations keep causing large-scale and long-lasting outages~\cite{meta-outage-2021,azure-outage-2023,cloudflare-outage-2025}.
Decades of research have answered with network verifiers that promise to catch bugs offline, before they reach production by comparing the configuration's \emph{behavior} against the operators' intent.
Yet, only a small fraction of production networks deploy them~\cite{krentsel2025towards}, and the reasons are well known in the literature: Verifiers \linebreak
\cModel~need high-fidelity models of deployed devices~\cite{brown2023lessons,krentsel2025towards,protogeros2026self}, \linebreak
\cTime~struggle to scale to large networks~\cite{tang2023lightyear,thijm2024kirigami,li2025ndd,schneider2025guided}, and \linebreak
\cFeedback~report misbehavior, but help little in debugging~\cite{tang2021campion,schneider2025guided}.

We agree with all three, but see them as symptoms of a single root cause: verifiers assume nothing about \emph{how} operators intend to implement the behavior.
Seeing only what \emph{is} configured, not what is \emph{intended}, they must be generic and support every possible implementation.
For instance, to reach external destinations, a network can propagate their routes over iBGP in various topologies, redistribute them into OSPF or IS-IS, or mix both with static routes on top.
A network deploys one of them, yet verifiers must accept them all: \linebreak
They must \cModel~model all protocols plus their interactions, \linebreak
\cTime~build the full symbolic network state before checking, and \linebreak
\cFeedback~cannot blame any command, as each could be intended.

We therefore argue that operators should state not only the intended behavior, but also \emph{how} they plan to implement it: the network's \emph{design}.
In the example above, the design would not require reachability, but that the iBGP sessions form a route-reflection topology.
Operators already make these design decisions, e.g., OSPF within regions and BGP in between, a dedicated VRF to isolate management traffic, and L3VPNs connecting customer sites.
A configuration \emph{implements} a design if each of its decisions hold; checking this \linebreak
\cModel~requires modeling only the commands the design uses, \linebreak
\cTime~amounts to a fast check on the configuration text, and \linebreak
\cFeedback~points directly at the invalid commands, like those configuring invalid iBGP sessions in the example above.

Hyperscalers seem to agree: their custom management systems~\cite{sung2016robotron,liu2018automatic,mogul2020experiences} make their design explicit, and either check configuration changes against it or generate them from the design.
But these systems are tailored to their own design and tightly integrated in their tooling; they would not fit other networks even if they were open-sourced.
The vast majority of networks lack the resources to engineer such a solution and are left with no tool to capture their design:\linebreak
Existing tools either check local configuration properties~\cite{beckett2026concord} or analyze the behavior produced by a configuration~\cite{feamster2005detecting,brown2023lessons}.
Although operators can script custom analyses over tools such as Batfish, those scripts are written against configurations or network state~\cite{batfish-docs}, not against an explicit representation of the network's design.

We instead propose \system, a language that makes the design itself a first-class citizen that can be checked.
It is built around two goals:
The first is to be \emph{flexible}: every network's design is different, so operators must be able to state their own decisions rather than pick from predefined checks.
\system therefore builds on Datalog: operators write design decisions as rules, and recursion lets those rules span the entire network rather than a single device.
The second is to operate \emph{directly on configurations}, not on a model derived from them.
A templating layer matches configuration commands and exposes their parameters to the design rules, linking each derived fact to the command it came from.
Thus, \system is tied to no vendor or device.
The same language can capture router configurations, and those from applications like DNS or web servers, letting operators express dependencies between applications and the network.
We find that writing a design is little work that pays off in fast checks and localized feedback:
140 lines suffice to check a 4-million-line configuration of a 754-node network, and \system does so in five seconds while locating the commands that violate the design.

\system's benefits are not for free.
With verifiers, the cost is complex models, slow checks, and vague errors; with \system, it is effort: operators must write the design and ensure it is correct.
We argue that both efforts are smaller than they appear.
Writing a design does involve modeling, but only proportionally: operators write templates only for the commands they use and rules only for the decisions they make, never describing how the network computes its forwarding state.
Further, they need not start from scratch: just as existing tools mine behavioral specifications~\cite{birkner2020configspec}, templates~\cite{kakarla2020selfstarter}, or device-local contracts~\cite{beckett2026concord} from configurations, future tools could mine the design a configuration already implements.

True to its name, \system takes the design as an axiom: any configuration that implements it is assumed correct.
If the design is incorrect, so is every conclusion built on it.
Yet, this assumption is nothing new, as operators today already trust their design.
That trust is implicit, and since only a few networks run a verifier, it goes entirely unchecked.
\system instead makes it explicit and at least guarantees that the configuration correctly implements the design.

Beyond checking, an explicit design opens opportunities for the tools that manage configurations.
A check that finishes in seconds and points at the offending command can run mid-edit and give network engineering the integrated development environment (IDE) that programmers have long enjoyed.
Further, a violated rule carries more than a location: its negative provenance~\cite{wu2014diagnosing} explains which facts are missing for the design to hold, the building blocks for automatically suggesting fixes.
The same feedback turns the checker into an oracle for LLMs that guarantees their edits preserve the design, not just the behavior~\cite{mondal2023llms}.
\system could even constrain the generation directly like type systems already do~\cite{mundler2025typeaware}.
Ultimately, provenance and constrained generation aim at the same goal: entire configurations produced and updated from the design itself, bringing the end-to-end tooling of hyperscalers to every other network.

%%% Local Variables:
%%% mode: LaTeX
%%% TeX-master: "main"
%%% End:

%% file: example.tex
\input{figures/language-overview}

\section{\system by Example}\label{sec:example}

Consider \Cref{fig:example}: the design (top) checks that two routers \Code{r1} and \Code{r2} (bottom) form an iBGP full mesh.
Like every \system design, it combines \Op{set} statements to \emph{model} the configuration, storing facts in a database, and \Op{assert} statements that \emph{check} properties over that database.
\system then generates a single network-wide Datalog program.

\paragraph{Local facts}
\Cref{lst:axiom-def-iface} matches every interface that configures an IP address, binding the interface name and the address.
It matches twice, and each executes the body with the bound values as constants: the \Op{set} statement inserts a fact into the \Rel{Addr} relation (declared on \Cref{lst:axiom-rel-addr}), yielding two Datalog facts: \Code{\Rel{Addr}(\Str{r1}, \Str{eth0}, 5.0.0.1)} and \Code{\Rel{Addr}(\Str{r2}, \Str{eth0}, 5.0.0.2)}.
\Cref{lst:axiom-def-bgp} stores all AS numbers in \Rel{Asn} the same way.

\paragraph{iBGP sessions}
Checking a BGP session requires more than the local configuration: \Code{r1} refers to its neighbor only by IP address, while that neighbor's name and AS number live in \Code{r2}'s configuration.
Datalog bridges this gap: the \Op{for} statement on \Cref{lst:axiom-for-ip-address} searches \Rel{Addr} for the router \Var{N} owning the configured IP---a string known only once the Datalog program runs.
Each statement in its body therefore compiles to Datalog joins with relation \Rel{Addr} over \Var{N}: the \Op{set} on \Cref{lst:axiom-set-ibgp-neighbor} becomes a \emph{rule} deriving the session in \Rel{IBgpSession}, and the \Op{assert} on \Cref{lst:axiom-assert-neighbor-asn} a \emph{query} requiring \Var{N} to have the configured \Code{remote\_asn}.
If no router owns the configured IP, no session is derived; the assertion on \Cref{lst:axiom-assert-ibgp-neighbor} catches exactly this case.

\paragraph{\normalfont\normalsize\color{black}\itshape iBGP full mesh}
\Rel{IBgpSession} contains all correctly configured sessions, so this check reduces to a single join with \Rel{Asn}: \Cref{lst:axiom-assert-full-mesh-1,lst:axiom-assert-full-mesh-2} assert that every router maintains a session with every other in its AS.
Note, that each \Op{assert} is a distinct failure mode: a mismatched \Code{remote-as}, a neighbor IP that no router owns, and a missing session.
If any fail, \system reports a user-defined message pointing to the offending command.

%%% Local Variables:
%%% mode: LaTeX
%%% TeX-master: "main"
%%% End:

%% file: figures/language-overview.tex
\tikzset{
  conn/.style={thick, line join=round, shorten <=-1pt, shorten >=-1pt},
}
\newcommand\rail[2]{\coordinate (#1) at #2;}
\newcommand\wirett[4][neutral400]{%
  \coordinate[yshift=+.35mm] (xx) at (#2.north);
  \coordinate[yshift=+.35mm] (yy) at (#3.north);
  \draw[conn, #1] (#2) -- (xx) -- (xx -| #4) -- (yy -| #4) -- (yy) -- (#3);%
}
\newcommand\wiretb[4][neutral400]{%
  \coordinate[yshift=+.35mm] (xx) at (#2.north);
  \coordinate[yshift=-.35mm] (yy) at (#3.south);
  \draw[conn, #1] (#2) -- (xx) -- (xx -| #4) -- (yy -| #4) -- (yy) -- (#3);%
}
\newcommand\wirebt[4][neutral400]{%
  \coordinate[yshift=-.35mm] (xx) at (#2.south);
  \coordinate[yshift=+.35mm] (yy) at (#3.north);
coordinate[yshift=-.35mm] (yy) at (#3.south);
  \draw[conn, #1] (#2) -- (xx) -- (xx -| #4) -- (yy -| #4) -- (yy) -- (#3);%
}
\newcommand\wirebb[4][neutral400]{%
  \coordinate[yshift=-.35mm] (xx) at (#2.south);
  \coordinate[yshift=-.35mm] (yy) at (#3.south);
  \draw[conn, #1] (#2) -- (xx) -- (xx -| #4) -- (yy -| #4) -- (yy) -- (#3);%
}

\begin{figure}%
  % highlight iface names
  \lmFit[fill=neutral200!70]{1}%
  \lmFit[fill=neutral200!70]{6}%
  \lmFit[fill=neutral200!70]{11}%
  % highlight IP addresses
  \lmfit[fill=blue100!70]{2}%
  \lmfit[fill=blue100!70]{7}%
  \lmfit[fill=blue100!70]{12}%
  % highlight ASN
  \lmfit[fill=purple100!70]{3}%
  \lmfit[fill=purple100!70]{8}%
  \lmfit[fill=purple100!70]{13}%
  % highlight neighbor IP
  \lmfit[fill=red100!70]{4}%
  \lmfit[fill=red100!70]{9}%
  \lmfit[fill=red100!70]{14}%
  % highlight remote asn IP
  \lmfit[fill=yellow100!70]{5}%
  \lmfit[fill=yellow100!70]{10}%
  \lmfit[fill=yellow100!70]{15}%
  % needed because the lmFit somehow creates vertical space.
  \vspace{-1.1mm}%
  \begin{lstlisting}
rel |Addr|(device: String, iface: String, addr: IpAddr);(*\label{lst:axiom-rel-addr}*)
rel |Asn|(device: String, asn: Int);(*\label{lst:axiom-rel-asn}*)
rel |IBgpSession|(device: String, neighbor: String, ip: IpAddr);(*\label{lst:axiom-rel-ibgp}*)

def ("interface", @iface: String@, "ip", "address", @ip: IpNet@):(*\label{lst:axiom-def-iface}*)
  set |Addr|(self, iface, ip as IpAddr); // Router <--> IP mapping

def ("router", "bgp", @asn: Int@) {(*\label{lst:axiom-def-bgp}*)
  set |Asn|(self, asn); // Router <--> AS number mapping(*\label{lst:axiom-set-asn}*)
  def ("neighbor", @ip: IpAddr@, "remote-as", @remote_asn: Int@) {(*\label{lst:axiom-def-neighbor}*)
    if asn == remote_asn {(*\label{lst:axiom-if-ibgp}*)
      for |Addr|(_ as $N$: String, _, ip) {(*\label{lst:axiom-for-ip-address}*)
        assert |Asn|($N$, remote_asn): "Invalid remote-as: " ++ $N$;(*\label{lst:axiom-assert-neighbor-asn}*)
        set |IBgpSession|(self, $N$, ip);(*\label{lst:axiom-set-ibgp-neighbor}*)
      }
      assert |IBgpSession|(self, _, ip): "No router has IP " ++ ip;(*\label{lst:axiom-assert-ibgp-neighbor}*)
    }
  }
  for |Asn|(_ as $N$: String, asn): if $N$ != self:(*\label{lst:axiom-assert-full-mesh-1}*)
    assert |IBgpSession|(self, $N$, _): "Missing session with " ++ $N$;(*\label{lst:axiom-assert-full-mesh-2}*)
}
\end{lstlisting}
  \begin{minipage}{0.5\linewidth}
    \begin{lstlisting}[language=Cisco, firstnumber=22]
! Configuration of r1
interface @eth0@
  ip address @5.0.0.1/24@
router bgp @5@
  neighbor @5.0.0.2@ remote-as @5@
\end{lstlisting}
  \end{minipage}%
  \begin{minipage}{0.5\linewidth}
    \begin{lstlisting}[language=Cisco, firstnumber=27]
! configuration of r2
interface @eth0@
  ip address @5.0.0.2/24@
router bgp @5@
  neighbor @5.0.0.1@ remote-as @5@
\end{lstlisting}
  \end{minipage}
  \caption{
    \system design to enforce an iBGP full-mesh, along with the configuration of two routers \Code{r1} and \Code{r2}.
  }
  \label{fig:example}
  \Description[Axiom design and two-router configuration example.]{
    The top listing defines relations for interface addresses, AS numbers,
    and iBGP sessions. Templates extract configuration facts, while assertions
    check neighbor AS numbers, resolve neighbor addresses to routers, and
    require an iBGP full mesh within each AS.
    Below, two side-by-side configurations show routers r1 and r2 in AS 5,
    with addresses 5.0.0.1/24 and 5.0.0.2/24 and reciprocal iBGP neighbors.
    Matching configuration values and template variables share colors.
  }
\end{figure}

%%% Local Variables:
%%% mode: LaTeX
%%% TeX-master: "../main"
%%% End:

%% file: motivation.tex
\section{Checking a Network Design}

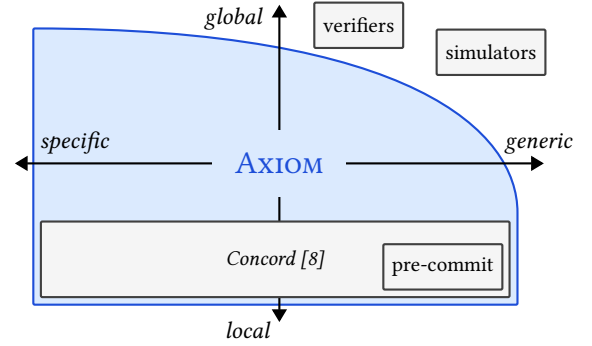
\begin{figure}
  \centering
  \begin{tikzpicture}[
    x=3.5cm, y=2.1cm,
    >=Triangle,
    axis label/.style={font=\small\itshape},
    existing/.style={
      draw, thick, rounded corners=.2mm,
      minimum height=6mm, align=center,
      font={\footnotesize}, fill=neutral100, draw=neutral700
    },
  ]
    \tikzmath{
      \xmin = -0.93; \xmax = 0.90;
      \ymin = -0.89; \ymax = 0.85;
    }

    \draw[fill=blue100, draw=blue700, thick, rounded corners=.2mm]
      (\xmin, \ymin) -- (\xmax, \ymin) -- ++(0, 12mm) .. controls (\xmax, 0.2) and (0.5, \ymax) .. (\xmin, \ymax) -- cycle;
  
    \draw[thick, <->] (-1,  0) node[axis label, above right, xshift=2mm] {specific}
                   -- ( 1,  0) node[axis label, above left, xshift=5mm] {generic};
    \draw[thick, <->] ( 0, -1) node[axis label, left, yshift=-1mm] {local}
                   -- ( 0,  1) node[axis label, below left, xshift=-1mm, yshift=1mm] {global};

    \node[existing]   at (+0.3, +0.87) {verifiers};
    \node[existing]   at (+0.8, +0.70) {simulators};
    % \node[existing, anchor=south east] at (\xmax, \ymin) {pre-commit\\checks};
    \node[existing, minimum width=62mm, minimum height=10mm, anchor=south west, outer sep=1mm] at (\xmin, \ymin) {\emph{Concord~\cite{beckett2026concord}}};
    \node[existing, anchor=south east, outer sep=2mm] at (\xmax, \ymin) {pre-commit};

    \node[blue700, inner sep=3mm, fill=blue100] at (0, 0) {\Large \system};
  \end{tikzpicture}
  \caption{\system enables writing designs ranging from specific to generic, and covering low-level and high-level properties.}
  \label{fig:relation}
  \Description[Two-dimensional map relating Axiom to existing network checkers.]{
    A Diagram with two axes crossing at the center.
    The horizontal axis ranges from specific on the left to generic on the right; the vertical axis ranges from local properties at the bottom to global properties at the top.
    Existing approaches appear as boxes: verifiers and simulators sit in the top right, checking global properties generically; pre-commit checks sit in the bottom right, checking local properties generically; Concord spans the bottom, checking local properties at any generality.
    A shaded region labeled Axiom covers most of the map: it spans the full horizontal range for local properties and extends up to global properties on the specific side, tapering off towards the generic and global corner.
  }%
\end{figure}

\Cref{fig:relation} relates \system to existing approaches along two dimensions.
The x-axis shows \emph{generality}: how many \emph{valid} configurations a checker accepts.
Verifiers and simulators are generic and (try to) accept any configuration with the intended behavior; a design accepts only those that
implement it.
The y-axis shows \emph{expressivity}: the scope of the checked properties, from local ones on a single device's configuration to global ones spanning the entire network.
Where a checker sits on these axes determines the cost of its checks and the precision of its errors, as we show in the following.

\paragraph{Generality}
Generic specifications need to be only stated once; they cover any configuration change, including ones that replace entire protocol stacks.
However, a specification is only tenuously connected to the configuration: proving that an arbitrary configuration satisfies it is expensive, and a violation cannot be attributed to a specific configuration edit.
A specific design inverts this trade-off: it is cheap to check and its errors point at concrete commands, but it rejects \emph{correct} configurations that implement a different design.

Crucially, \system can express both ends of this trade-off, as two designs for iBGP route propagation illustrate.
A generic design states only the intended behavior: a route originated at any router must reach every other router.
\system can express this through recursion on the propagation graph and an assertion that tests every router pair.
Doing so, however, is expensive as our evaluation in \cref{sec:evaluation} shows, and if a session is missing, it reports that routes fail to propagate, not \emph{why}.
The full-mesh design of \Cref{sec:example} sits at the specific end: it fixes one implementation.
The checks reduce to simple joins, and a missing session can be flagged directly.

\paragraph{Expressivity}
Existing tools sit at the two ends of this axis: pre-commit checks and configuration contracts~\cite{beckett2026concord} check local properties of a single configuration, such as whether every referenced route-map exists, while verifiers and simulators check global ones over the converged routing and forwarding state.
Local properties are natural to express in \system; the question is whether a design can capture global ones \emph{without} computing that state.

We find that it can, again by fixing the implementation.
Consider a waypoint property: traffic towards a server must traverse a firewall.
A verifier must support every possible implementation, but a design fixes one.
Say the network implements the waypoint with IPv6 segment routing.
The property then holds if \emph{(i)}~the IGP is configured correctly, \emph{(ii)}~every border router inserts a firewall segment, and \emph{(iii)}~no static route overrides the IGP towards the firewall or the server---three checks directly on the configuration text.
Notably, this guarantee extends to every failure in which the network stays connected, without any explicit failure model.

Beyond correctness, a design can also enforce performance objectives.
To respect a hardware limit on the forwarding table size, for example, a design may require strict filters on incoming BGP routes, a cap on routes propagated in OSPF, and a default route.
Where the limits of this approach lie remains open, but we conjecture that most reachability and security properties can be expressed as a design by trading generality for expressivity.

%%% Local Variables:
%%% mode: LaTeX
%%% TeX-master: "main"
%%% End:

%% file: language.tex
\section{\system}

\Cref{sec:example} introduced \system's language by example.
This section presents the machinery around it: the inputs, how configurations and designs map to Datalog, and how our prototype executes them.

The main input of \system is a \networktoml.
It contains the network topology, that is, all devices, their interface names, and all connections.
\system automatically populates some Datalog relations with this information, making it accessible from the design.
Besides the topology, for every device, it lists the path to its configuration files and how to parse them, and the \extension files containing the design to check.

\system parses each configuration into a common tree-like structure.
This tree is \system's only assumption about its input: whether the configuration belongs to a router, a switch, or an application like a DNS server, the same design language can match it, and express dependencies across all of them.
Nested \Op{def} statements (\Cref{lst:axiom-def-iface,lst:axiom-def-bgp,lst:axiom-def-neighbor}) match paths in this tree, and their match types (\Ty{String} or \Ty{IpNet}) determine how the configuration values are interpreted.

While \system unifies the configuration syntax into a tree, it does not unify the semantics; doing so would require modeling every vendor and every subtle difference between their devices.
Instead, the semantics live in the design itself, which remains vendor-specific.
To avoid duplication, the \Op{apply} mechanism shares parts of a design across vendors.

\subsection{Stratified Datalog with Lattices}
Unlike most verifiers, \system relies on Datalog instead of a general-purpose solver like SMT.
We do so for two reasons.
First, Datalog's cost is predictable: Operators can identify the rules that dominate execution time or memory, and improve them by reordering joins or adding guards.
Second, Datalog derives all facts, not just a verdict.
While this may seem wasteful, every derivation carries provenance~\cite{wu2014diagnosing} leading to configuration lines.
This lets tooling explain why a design fails and what commands are responsible, or even suggest completions.

Datalog also matches the shape of the problem.
Protocol semantics and the topology relate configuration objects across devices (like how the BGP neighbor statement of one router relates to an IP address of another in \Cref{sec:example}) and a join follows such a relation.
Global properties, however, must follow such relations transitively, which requires recursion.
It is exactly this lack of recursion that prevents \emph{Concord}'s contracts~\cite{beckett2026concord} from expressing global properties.

Recursion alone is not enough, as designs also constrain what must be \emph{absent}: an interface that is shut down, for example,
disables all its configuration.
Datalog only derives facts and never retracts them, so a rule may check for the absence of a tuple only once that relation is fully evaluated.
\emph{Stratification} orders rules to guarantee any computed fixpoint is consistent.

A common operation of routers and switches is to reduce many candidates to a single best one, e.g., the longest prefix, the shortest path; an operation off-the-shelf Datalog does not support.
Recent Datalog variants model such a column as a \emph{lattice}~\cite{madsen2016datalog,arntzenius2016datafun,madsen2020fixpoints} that imposes an ordering on the column's values, and the engine keeps only the best one.
A consistent fixpoint requires rules on lattices to be \emph{monotone}, that is, to only grow the value; we resolve non-monotone rules with stratification.

Recursion, stratification, and lattices are the ingredients needed to express every design we have written thus far, including generic ones that compute parts of the routing tables.
The language itself is still young: it can encode prefix-lists and route-maps, but not always conveniently, and a few features are still missing entirely (discussed in \Cref{sec:discussion}).

\subsection{Execution}

We implement a prototype of \system in Rust\footnote{\url{https://github.com/nsg-ethz/axiom}}.
It runs in three steps: parsing all configuration and \extension files, evaluating the resulting intermediate representation (IR), and running the generated Datalog program.
Configurations are parsed as specified in the \networktoml.
For \extension files, we specify its grammar as a PEG~\cite{ford2004peg}.
The remaining steps are detailed as follows:

\paragraph{Evaluating the IR}
This step matches all \Op{def} blocks against the parsed configurations to generate a single Datalog program.
\system recursively walks each configuration tree while maintaining, for every matching \Op{def} block, an evaluation context that binds the matched variables and accumulates all \Op{if} conditions and \Op{for} queries.
From this context, it generates the rules and queries that constitute the Datalog program.
Two kinds of variables appear along the way: those bound to concrete configuration values are evaluated directly in the IR, while symbolic ones from \Op{for} bindings (capitalized and colored red in \Cref{fig:example}) are left to the Datalog engine.

\paragraph{Datalog engine}
We implement our own Datalog engine to support lattice semantics.
A lightweight \emph{motion analysis} first tracks how each value evolves during the fixpoint computation: it only grows, only shrinks, remains stable, or is none of the three.
A rule is monotone if its values are all stable, or if the lattice agrees with its motion, e.g., a minimum lattice accepts only stable or shrinking values.
This analysis stratifies the program, akin to monotonic aggregation~\cite{ross1992monotonic}.

A simple query planner prepares table indices from the join patterns of the rules: non-lattice relations are indexed whenever a join needs it, whereas lattice relations are indexed once fully evaluated.
\system then executes each stratum using semi-na\"ive evaluation~\cite{bancilhon1986naive}, and checks all \Op{assert}s once their referenced relations are fully evaluated.

Our prototype does not yet evaluate incrementally, but the literature on incremental Datalog is extensive~\cite{gupta1993maintaining,motik2015incremental,hu2018optimised,ryzhyk2019datalog}, including for lattice semantics~\cite{szabo2018incrementalizing,szabo2021incremental}.
We see an incremental engine as a promising direction, especially for checking the configuration on every keystroke.

%%% Local Variables:
%%% mode: LaTeX
%%% TeX-master: "main"
%%% End:

%% file: evaluation.tex
\section{Early Results}\label{sec:evaluation}

In the following, we investigate three questions about \system:
\begin{enumerate}[leftmargin=*, label={\itshape{(\roman*)}}]
\item whether its language captures designs concisely,
\item whether its checker scales to large networks, and
\item how specific designs differ from generic ones.
\end{enumerate}

To answer them, we automatically generate Cisco configurations for all networks in TopologyZoo~\cite{knight2011internet}, the largest containing 754 routers.
The configurations cover interface IP addresses, OSPF with one or multiple areas and virtual links, and iBGP sessions with either a full mesh or route reflection.
We then write a generic \system design that accepts all combinations, and specific ones that pin down which variant is used and how.
We run the experiments on an AMD EPYC server using 16 threads.

\paragraph{The language of \system is concise and expressive}
We split the design into three files, covering interfaces, OSPF, and iBGP, and write the latter two in a specific and a generic variant.
The interface design relates devices: addresses must be unique, and connected interfaces must share a subnet.
The specific OSPF design requires each area to be internally connected (a recursive property) and adjacent to the backbone.
Further, it requires every interface address to be advertised.
These two properties then establish reachability among all interface addresses.
The specific iBGP designs rely on this reachability to guarantee all iBGP sessions can be established.
We write two versions: one that requires a full mesh, resembling the example from \cref{sec:example}, and one where only the route reflectors need to form a full mesh, and all others have iBGP sessions with at least one reflector.
The resulting \extension files need only 140 lines to check configurations of over 4 million lines.
Of these, 19 templates and 18 rules model every command the designs use.

The generic variant relaxes requirements for both OSPF and iBGP.
OSPF no longer needs to advertise every interface; instead, the design propagates each network as far as the configuration allows.
Whether two routers can establish an iBGP session now depends on whether they receive the other's network.
Instead of pinning down a specific topology, the iBGP design builds the propagation graph and asserts a valid path between all pairs of routers.
Despite the added complexity, the design files have only 162 lines in total, and needs 8 additional rules compared to the specific design.

\begin{figure}
  \centering
  \input{figures/time-eval}
  \caption{
    \system scales to large networks, checking their configurations within seconds.
    An iBGP full mesh needs almost 100 times more commands than route reflection, explaining the difference in checking time.
  }
  \label{fig:time-eval}
  \Description[Log-log plot of checking time versus network size]{
    A log-log line plot showing \system's running time against the number of routers, from 20 to 754, for two iBGP variants.
    Both curves grow steadily; checking route reflection takes below 0.3 seconds, while checking a full mesh takes at most 3 seconds across all networks.
  }
\end{figure}
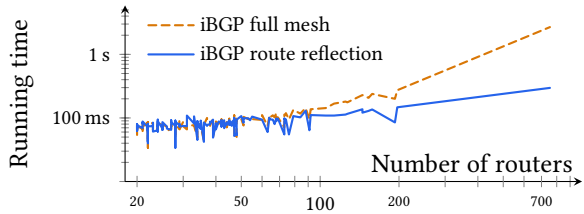

\paragraph{\system can scale to large networks}

\Cref{fig:time-eval} shows the running time of the specific designs with four OSPF areas on all networks in TopologyZoo.
With route reflection, \mbox{\system} checks every network within 0.3 seconds, including the largest with 754 routers.
The full mesh takes up to five seconds, but the cause is the input itself: a session between every pair of routers grows the largest configuration to 4.57 million lines, almost 100 times more than with route reflection.
This is a stress test rather than a deployment; an iBGP full mesh across 754 routers does not scale~\cite{rfc4456}.
As a comparison, \emph{Batfish}~\cite{fogel2015general} spends 12 minutes parsing the configurations and 3 simulating the network---more than two orders of magnitude longer than \system.
This comparison illustrates the cost of computing network behavior, which \system's approach avoids by checking the design directly.

\paragraph{Specific designs yield cheaper checks and localized errors}

\begin{table}
  \footnotesize
  \begin{tabular}{lrlrr}
    %\toprule
    Configuration                & \# lines & Design   & Time    & Memory   \\\midrule
    Single area, iBGP full mesh  & 4.58\,M  & generic  & 4.67\,s & 6.35\,GB \\
                                 &          & specific & 2.64\,s & 4.19\,GB \\[1mm]
    Four areas, iBGP full mesh   & 4.57\,M  & generic  & 5.46\,s & 6.65\,GB \\
                                 &          & specific & 2.69\,s & 4.13\,GB \\[1mm]
    Four areas, route reflection & 0.05\,M  & generic  & 2.27\,s & 2.53\,GB \\
                                 &          & specific & 0.29\,s & 0.27\,GB \\
    %\bottomrule
  \end{tabular}
  \caption{
    Checking specific designs is easier than generic ones.
    We compare the running time and memory usage of \system to check specific or generic designs on the largest network in our dataset.
  }
  \label{tab:generality}
\end{table}

\Cref{tab:generality} compares generic and specific designs on the largest network.
The specific designs are cheaper to check: almost 2$\times$ for the iBGP full mesh, and 10$\times$ for route reflection, with a tenth of the memory.
Derived facts drive memory usage: the generic designs require facts for OSPF routes and propagation paths, and the full mesh millions of session facts.

With the specific design, \system accurately locates the errors; with the generic one, it reports only the high-level symptom, not the responsible command.
When OSPF does not advertise an address needed by iBGP, the specific design reports the exact interface that is missing from OSPF, while the generic one reports that the IP address cannot be reached.
When a full mesh misses a session, the specific design tells exactly which one, while the generic one warns that routes from either endpoint do not reach the other.
Note that a verifier would report both errors as the same misbehavior: externally advertised routes would not be reachable.

%%% Local Variables:
%%% mode: LaTeX
%%% TeX-master: "main"
%%% End:

%% file: figures/time-eval.tex
\begin{tikzpicture}
  \begin{axis} [
      axis lines=left,
      width=.9\linewidth,
      height=3.915cm,
      xmode=log,
      ymode=log,
      xmin=18, xmax=1000,
      ymin=0.01, ymax=5.9,
      xlabel={Number of routers},
      ylabel={Running time},
      tick label style={font=\footnotesize},
      xtick={100, 900},
      xticklabels={100,},
      extra x ticks={20,30,40,50,60,70,80,90,200,300,400,500,600,700,800},
      extra x tick labels={20,,,50,,,,,200,,,,,700,,},
      extra x tick style={tick label style={font=\tiny},},
      ytick={0.01, 0.1, 1, 10},
      yticklabels={, 100\,ms, 1\,s, 10\,s},
      every axis x label/.append style={at={(axis description cs:1.0, 0.0)}, anchor=south east},
      legend pos=north west,
      legend style={font=\footnotesize,fill=none,draw=none, inner ysep=0pt, outer ysep=0pt},
      legend cell align=left,
    ]

    \addplot[thick, amber600, densely dashed, line cap=round, line join=round, keep if all three={
      ospf_config is 3-areas and
      bgp_config is full-mesh and
      annotation_style is specialized
    }] table [col sep=comma, x=n_routers, y expr=\thisrow{elapsed_ms}/1000] {figures/results-new.csv};
    \addlegendentry{iBGP full mesh}

    \addplot[thick, blue600, solid, line cap=round, line join=round, keep if all three={
      ospf_config is 3-areas and
      bgp_config is rr-k5-r2 and
      annotation_style is specialized
    }] table [col sep=comma, x=n_routers, y expr=\thisrow{elapsed_ms}/1000] {figures/results-new.csv};
    \addlegendentry{iBGP route reflection}
  \end{axis}
\end{tikzpicture}

%%% Local Variables:
%%% mode: LaTeX
%%% TeX-master: "../main"
%%% End:

%% file: related-work.tex
\section{Related Work}
\system relates to three lines of work: tools that, like \system, check the configuration itself; tools that check the behavior it produces; and programming languages to capture network computation or behavior.

\paragraph{Query-based configuration checking}
\emph{Rcc}~\cite{feamster2005detecting} normalizes configurations into a database and queries it for predefined, known BGP faults.
\emph{SelfStarter}~\cite{kakarla2020selfstarter} flags outliers among similar configuration segments; it checks consistency, not intent.
Further, it cannot relate one device's configuration to another's, say, a BGP neighbor command to the router owning that address.
\emph{NCGuard}~\cite{vanbever2008ncguard} and \emph{Concord}~\cite{beckett2026concord} instead let operators write their own conditions:
\emph{NCGuard} as queries over an XML representation of the configuration, and \emph{Concord} as contracts attached to commands.
Yet, both fail to capture network designs that involve properties on a graph as they lack recursive queries.
\system checks everything these tools can, plus the recursive properties they cannot.

\paragraph{Behavioral verification and synthesis}
Simulation~\cite{quoitin2005modeling,fogel2015general,brown2023lessons,schneider2025extending} computes the behavior of a configuration, and data-plane~\cite{kazemian2012header,khurshid2012veriflow,guo2022flash} and control-plane verification~\cite{fayaz2016efficient,gember2016fast,beckett2017general,tang2023lightyear,wang2024expresso,schneider2025guided,schneider2025verifying} check that behavior against the operators' intent.
Synthesis~\cite{beckett2017network,el2018netcomplete,abhashkumar2020aed,schneider2022complexity} inverts the direction and generates configurations from the intended behavior.
LLM-based management~\cite{mondal2023llms,zhou2024large,wang2024netconfeval,han2025network} edits configurations directly and regains guarantees only by invoking a verifier in a loop.
All these tools reason about what the network should do, but not how the configuration should achieve it.
Yet, each keeps its place beside \system: simulation and verification can validate the design itself, and synthesis and LLMs can write configurations that \system holds to the design.

\paragraph{Programming languages for networks}
Propane~\cite{beckett2016propane} compiles routing policy into BGP configurations, leaving the design to its compiler.
NetKAT~\cite{anderson2014netkat,foster2016probabilistic,moeller2024katch} programs and verifies the data plane, never touching the configuration.
\emph{NDlog}~\cite{loo2009declarative} may appear closest to \system, as both extend Datalog; yet, it models the route computation itself and derives the routing state as its fixpoint.
Even the semantics differ: \emph{NDlog} expires sub-optimal paths by attaching timers to facts, while \system keeps fixpoints well-defined with lattices and stratification, which even our generic designs never exhaust.
All three capture the network's behavior or its computation; \system is their counterpart for its design.

%%% Local Variables:
%%% mode: LaTeX
%%% TeX-master: "main"
%%% End:

%% file: discussion.tex
\section{Discussion and Outlook}\label{sec:discussion}
We close by discussing what the language still misses and what it can borrow from declarative networking.

\paragraph{Missing language features}
Writing designs revealed two gaps in the language.
First, configurations contain ordered lists of filters and transformations, like route-maps, where the first matching clause wins.
Expressing them today takes a lattice to recover the order and recursive rules to walk it.
We plan to instead let operators map such configuration objects to custom functions that the checker can call.

Second, \system cannot yet interpret regular expressions embedded in configurations, like AS-path filters.
Vendors define their own dialects, so the language must let operators define a dialect and then apply it.
Evaluating a regex on concrete strings is then simple, as Datalog derives facts only forward.
Some designs, however, need more: whether a filter drops every route with AS 666 in its path is a question about the regex itself, answered by inspecting its state machine rather than by testing strings.

\paragraph{Negative provenance and synthesis}
\emph{Y!}~\cite{wu2014diagnosing} extends \emph{NDlog} with \emph{negative provenance}: it explains a missing route by the chain of facts that would have derived it.
The technique transfers directly to \system, where a violated assert asks exactly that question: which facts are missing to correctly implement the design.
We expect the answers to improve as well.
In \emph{NDlog}, almost any missing route can be forced with a static route, so explanations range over every configuration that could produce it, intended or not.
A design admits only the intended mechanism, narrowing the explanation to, e.g., the missing iBGP session from \Cref{sec:evaluation}.

Negative provenance is the first step toward synthesis.
Turning the missing facts into suggested commands requires reversing our templates: a \Op{def} that matches commands must instead generate them.
Applied repeatedly, such suggestions could complete a configuration from the design alone.

%%% Local Variables:
%%% mode: LaTeX
%%% TeX-master: "main"
%%% End: